\documentclass[useAMS,usenatbib]{mn2e}
\usepackage{epsfig}

\def\lesssim{\mathrel{\hbox{\rlap{\hbox{\lower4pt\hbox{$\sim$}}}\hbox{$<$}}}}
\def\gtrsim{\mathrel{\hbox{\rlap{\hbox{\lower4pt\hbox{$\sim$}}}\hbox{$>$}}}}
\def\la{\mathrel{\hbox{\rlap{\hbox{\lower4pt\hbox{$\sim$}}}\hbox{$<$}}}}
\def\ga{\mathrel{\hbox{\rlap{\hbox{\lower4pt\hbox{$\sim$}}}\hbox{$>$}}}}

\def\spose#1{\hbox to 0pt{#1\hss}}
\def\approxlt{\mathrel{\spose{\lower 3pt\hbox{$\sim$}}
	\raise 2.0pt\hbox{$<$}}}
\def\approxgt{\mathrel{\spose{\lower 3pt\hbox{$\sim$}}
	\raise 2.0pt\hbox{$>$}}}

\def\<{\thinspace}

\def\boxit#1{\vbox{\hrule\hbox{\vrule\kern3pt\vbox{\kern3pt
          #1 \kern3pt}\kern3pt\vrule}\hrule}}

\def\ga{{\rm\thinspace gauss}}

\def\h50{\hbox{$\rm\thinspace h_{50}$}}
\def\h50m1{\hbox{$\rm\thinspace h_{50}^{-1}$}}

\title[Globular Clusters in the outer halo of M31]
{Globular clusters in the outer halo of M31: the survey}
\author[Huxor et al.]{A. P. Huxor$^{1}$,  N. R. Tanvir$^{2}$, A. M. N. Ferguson$^{1}$, M. J. Irwin$^{3}$, R. Ibata$^{4}$,
\newauthor   T. Bridges$^{5}$, G. F. Lewis$^{6}$\\
$^{1}$Institute for Astronomy, University of Edinburgh, Royal Observatory, Blackford Hill, Edinburgh EH9 3HJ\\
$^{2}$Department of Physics and Astronomy, University of Leicester, University Road, Leicester LE1 7RH
\\
$^{3}$Institute of Astronomy, Madingley Road, Cambridge, CB3 0HA \\
$^{4}$Observatoire de Strasbourg, 11, rue de l'Universite, F-67000, Strasbourg, France \\
$^{5}$Department of Physics, Queen's University, Kingston, Ontario, Canada K7M 3N6 \\
$^{6}$Institute of Astronomy, School of Physics, A29, University of Sydney, NSW 2006, Australia}

\begin{document}

\date{}

\pagerange{\pageref{firstpage}--\pageref{lastpage}} \pubyear{2007}

\maketitle

\label{firstpage}

\begin{abstract}

  We report the discovery of 40 new globular clusters (GCs) that have
  been found in surveys of the halo of M31 based on INT/WFC and
  CHFT/Megacam imagery. A subset of these these new GCs are of an
  extended, diffuse nature, and include those already found in \citet{Huxoretal05}. 
  The search strategy is described and basic
  positional and V and I photometric data are presented for each cluster. 
  For a subset of these clusters, K-band photometry is also
  given. The new clusters continue to be found to the limit of the survey area ($\sim$ 100 kpc),
  revealing that the GC system of M31 is much more extended than
  previously realised. The new clusters increase the total number of
  confirmed GCs in M31 by approximately 10\% and the number of
  confirmed GCs beyond 1$\degr$ ($\approx 14$~kpc) by more than 75\%.
  We have also used the survey imagery as well recent HST archival
  data to update the Revised Bologna Catalogue (RBC) of M31 globular
  clusters.

\end{abstract}


\section{Introduction}

Globular cluster (GC) systems provide fossil signatures with which to
untangle the history of galaxy formation. GCs are gravitationally
bound concentrations of (typically) between 10$^4$ to 10$^6$ stars, and it is
believed that they were among the first objects to be formed in a
galaxy, and so provide insight into the formative stages of the host
(see the review by \citealt{BrodieStrader06}). There is also strong
evidence that they form in major galaxy interactions and mergers, and
so can trace such events (e.g. Whitmore \& Schweizer 1995). Since GCs most
are believed to comprise stellar populations of a single age and
abundance (some unusual GCs, such as $\omega$ Cen and NGC
2808 exhibit multiple populations -
\citealt{Bedinetal04,Piottoetal07}), their colour-magnitude diagrams
(CMDs) provide a particularly powerful tool with which to age-date
significant events in a galaxy's history.  Indeed, the presence of
various sub-populations within a GC system is widely interpreted as
indicating distinct epochs of mass accretion and/or major star
formation (West et al. 2004).

In the case of the Milky Way, accretion events appear to have
played an important role in shaping the GC system.  Perhaps as many as eight of
the Milky Way halo GCs beyond a galactocentric radius R$_{GC}\gtrsim
10$~kpc ($\sim 20$\% of the population) are associated with the
Sagittarius dwarf spheroidal and its tidal stream
\citep{Bellazzinietal03, Sbordoneetal05, Tautvaisienetal04}, attesting
to the fact that even relatively minor accretion events can
significantly augment the globular cluster population of the parent
galaxy.  A further group of GCs and open clusters has been identified
as having a possible origin in the putative dwarf galaxy remnant in
Canis Major \citep{Martinetal04}, although at least one of these (NGC
2808) has since been shown to not be a member, from its orbit
\citep{Casettietal07}. Spatial alignments of Oosterhoff subpopulations
amongst other outer Milky Way GCs suggest that many more may have been
captured from accreted dwarfs (e.g. Yoon \& Lee 2002). It would be
valuable to discover if other galaxies have evidence of comparable GC
accretion.

The only large GC systems (more than 100 clusters) whose members can
be cleanly resolved into stars (and hence plotted on CMDs) with
current technology belong to the Milky Way (MW) and M31.  The GC
system of M31 is of particular interest since it allows for a direct
comparison between ages and metallicities derived from resolved
stellar population studies with HST, and from studies of integrated
light.  Moreover, as the nearest large spiral, M31 provides a means to
test how representative the Milky Way's GC system is of large spirals
in general.

\cite{BrodieStrader06} have summarised some of the major properties of
the M31 GC system, as part of their larger review on extragalactic GC
systems.  They conclude that the GC system of M31 to be quite substantial,
estimated at about 450 $\pm$ 100 members, which is a factor of $\sim$
3 greater than that of our own Galaxy.  To a first approximation, the
M31 GC luminosity distribution is similar to that of the MW in being
well-approximated by a Gaussian, although the details of the turnover
magnitude and dispersion may differ somewhat.  The M31 GC system also
contains a metal-rich and a metal-poor component, which are
tentatively identified as analogues to the MW disk-halo (or more
recently bulge-halo) GC populations \citep{Barmbyetal00}.  Early work by
\citet{vandenBergh69} hinted at a bimodal metallicity distribution in the M31 GC population.
Spectroscopy by \citet{Perrettetal02} confirmed this bimodality, with the peaks at [Fe/H] =
-1.44 and -0.50 dex.  Of their 301 GCs with spectroscopic
metallicities, they find 70 (i.e. 23\%) to be attributable to the
metal-rich population.  \citet{Perrettetal02} also present evidence
for a slight metallicity gradient in the metal-poor population, which,
although consistent with a scenario of early dissipational collapse,
does not preclude other models. There is also a correlation between the
 rotation of the GC populations and metallicity, first identified by \citet{vandenBergh69}.
More recently  \citet{Perrettetal02} have further shown that the
metal-rich GCs are centrally concentrated, with significant rotation
(160 km s$^{-1}$), and are consistent with a bulge population. The
metal-poor cluster system is more spatially extended, as expected for
a halo population, and intriguingly also exhibits rotation (131 km
s$^{-1}$).  
These results While the MW and the M31 GC systems share many common
properties, recent work has suggested that M31, unlike the MW, may
host a population of young GCs
\citep{Puziaetal05,Beasleyetal04,Bursteinetal04,FusiPeccietal05}.  If
confirmed, such a population could place interesting constraints on
the merger history of M31.

To date, the study of M31 GCs has been largely based on the excellent
Bologna Catalogue, which is frequently revised. Updates on the
original Revised Bologna Catalogue (RBC) of M31 GCs
\citep{Galletietal04a} were recently published \citep{Galletietal06,
  Galletietal07}. The RBC consists of the original Bologna Catalogue
\citep{Battistinietal87} supplemented with candidates from
\citet{Battistinietal93}, \citet{Auriereetal92},
\citet{Mochejskaetal98} and the HST archive work of
\citet{BarmbyHuchra01}.  Since it is based on the original Bologna
Catalogue \citep{Battistinietal87} which is taken from a 3 x 3 square
degree region around the centre of M31, the Revised Catalogue is also
mainly concentrated within these limits, i.e. a projected
galactocentric distance less than about 30 kpc (throughout this paper we assume 
a distance to M31 of $\sim$780 kpc,
\citealt{McConnachieetal05}).  However we find
globular clusters at much greater galactocentric distances (e.g. AM 1
at more than 120 kpc) from the centre of the MW, the most comparable
galaxy, suggesting that many undiscovered GCs may lie beyond the
Catalogue limits.  Many of the RBC candidates have been selected from
previous surveys based on photographic plates and contamination has
proved a significant problem. For example, \citet{Racine91} found from
high-resolution CFHT imaging that of 107 Bologna Catalogue candidates,
only 51 were genuine globular clusters. This is in spite of the fact
that he only chose candidates well away from the main disk of M31 to
avoid problems of crowding, extinction and reddening.  Spectroscopic
(e.g. \citealt{Barmbyetal00,Perrettetal02}) and imaging confirmation
of many candidates has been undertaken for the main disk region, and
along the major and minor axes, but little has been done for other
regions, except for serendipitous HST imaging as part of a programme
looking at background field galaxies \citep{BarmbyHuchra01}.

We present here a new search for GCs in the halo of M31, based on new
deep, wide-field imaging surveys undertaken at the INT and CFHT. In
the past few years, our group has been conducting a long-term study of
M31, using both ground-based and space-based instruments. These
surveys have led to the the discovery of many significant features in
M31 including copious low surface brightness substructure
\citep{Ibataetal01, Fergusonetal05, Ibataetal07}, an extended metal
poor stellar halo \citep{Chapmanetal06}, a giant rotating structure
\citep{Ibataetal05} and several new dwarf galaxies
\citep{Chapmanetal05, Martinetal06, Ibataetal07}.  Early results from
the search for new M31 GCs from these data were published in
\citet{Huxoretal04,Huxoretal05}.

\section{The Search for M31 clusters}

\subsection{The data}

The images and catalogues employed in this study were taken from two
different ground-based surveys.  The majority of the data were taken
as part of the Isaac Newton Wide Field Camera survey of M31 conducted
during the period 2000-2005. Exposures of 800-1000 seconds were taken
in the Johnson V and Gunn i bands, reaching (average 5 $\sigma$)
limiting magnitudes of i = 23.5 and V =24.5.  The average seeing was
generally better than 1.2\arcsec.  The fields observed include an area
far into the halo, and an additional region south along the Andromeda
Stream \citep{Ibataetal95} towards M33. The survey data were processed by the INT WFS
pipeline \citep{IrwinLewis01} provided by the Cambridge Astronomical Survey Unit; this
pipeline provides astrometry, photometry and object description and
classification (i.e. whether the object is stellar, non-stellar or noise).
 The INT/WFC fields that cover the
bright part of the M31 disk were not used in the search for new GCs
due to the high degree of crowding present in these regions. 
  The second dataset was taken from out still ongoing
Canada France Hawaii Telescope survey of the M31 far outer halo using
Megacam. Exposures of $5\times290$ seconds were taken in each of the g
and i bands with seeing typically better than 0.8\arcsec.  Full
details of these surveys can be found in \citet{Fergusonetal02} for
the INT/WFC data, and in \citet{Ibataetal07} for the CFHT/Megacam
data. We utilize the full INT WFC survey area corresponding to 60
square degrees but only 24 square degrees of the Megacam survey area.
This is due to the fact that only part of the latter dataset was in
hand at the time that our GC search was undertaken.

\subsection{The Strategy for Classical GCs}

In the Milky Way, star clusters are well resolved so searches for new
GCs can be based solely on star count density enhancements
(e.g. \citealt{Drake05}).  However, for most external galaxies, GCs
are unresolved and hence other search methods must be utilised. The
most common approach to identifying extragalactic GC candidates is to
use magnitude and colour information, as GCs have magnitudes and
colours that are expected to fit within a limited range.
Additionally, shape parameters such as size and ellipticity can also
be used. An example of this approach is provided by
\citet{Sharinaetal05} in their search for GC candidates in nearby (2 -
6 Mpc) dwarf galaxies.  Typically these searches adopt criteria based
on MW GC templates and thus have the inherent weakness of not probing
other regions of parameter space. That is, our view of what
constitutes a globular cluster will be biased towards objects that are
similar to the GCs found in our own Galaxy.

In very nearby galaxies, GCs are expected to be more extended than the
stellar PSF. \citet{Kimetal02}, for example, use the two radial
moments $r_{1}$ and $r_{-2}$ which characterize the image wing spread
and the image central concentration respectively.  A variant of this
approach has been recently employed by \citet{Gomezetal06} in their
search for GCs in NGC 5128. They subtract the stellar PSF from all the
sources in their frames, and visually inspect the residuals. Extended
objects leave a ``doughnut-shape'', as they are
under-subtracted in the wings and over-subtracted in the centre.
Similarly, \citet{Galletietal04b} use this technique to locate GCs in
NGC 253. However, PSF subtraction will also reveal compact background
galaxies as well as GCs hence additional rejection methods must be
employed.

In more distant galaxies, the only method available is to search for a
statistical over-density of objects with the characteristics of GCs
\citep{vandenBerghHarris82}.  This places constraints on the overall
number density and spatial extent of a GC system without being certain
of the status of any specific object. Control fields are usually
observed to obtain a background count, or alternatively, published
counts for likely contaminants, such as galaxies, can be used
\citep{LillerAlcaino83}.  One of the early applications of this
approach was in the hunt for M31 GCs \citep{Wirthetal85}.

Extragalactic GC candidates usually require spectroscopic
confirmation, where possible, since the risk of false positives is
high with the techniques discussed above. For example,
\citet{BeasleySharples00} undertook spectroscopic follow-up of 103
published GC candidates in the Sculptor group galaxies NGC 253 and NGC
55. The candidates had been selected on the basis of colour, magnitude
and galactocentric distance ($R_{gc}$) cuts, however, only 14 genuine
GCs in NGC253 and one probable GC in NGC 55 were actually confirmed.

\begin{figure}
 \centering
 \includegraphics[width=80mm]{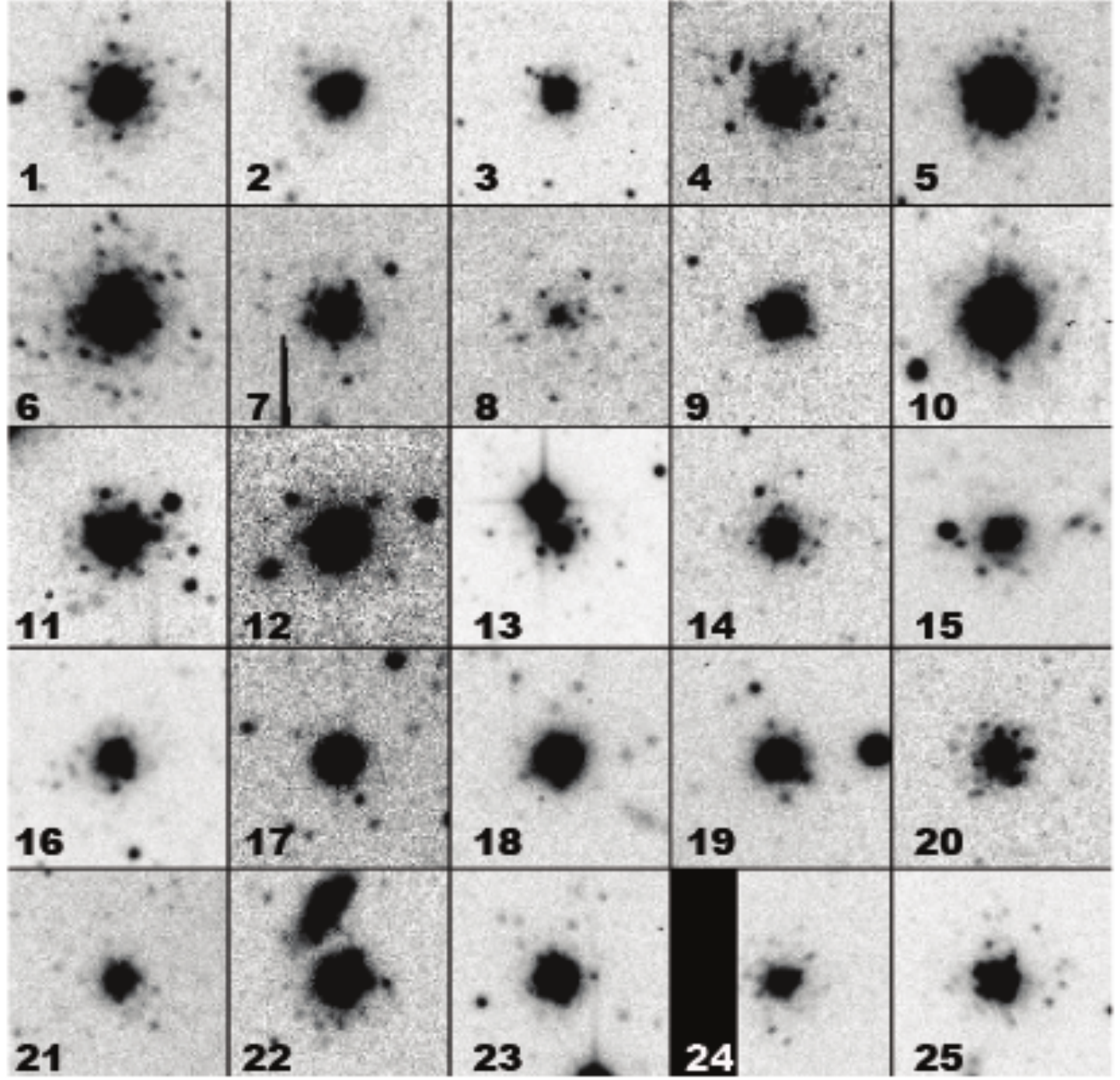}
 \vspace{2pt}
 \caption[V-band images of the new clusters from the INT data]
 {V-band images of the new clusters from the INT data.  Each image is
   30 x 30 arcsec.  They are in RA order (from H1 to H25) left to
   right and top down.  One cluster is a partial image as it lies on
   the edge of the chip, and one (in the centre of the mosaic)
   overlaps the image of a nearby bright
   star.}\label{Fi:classical-int}
\end{figure}

\begin{figure}
 \centering
 \includegraphics[angle=0,width=35mm]{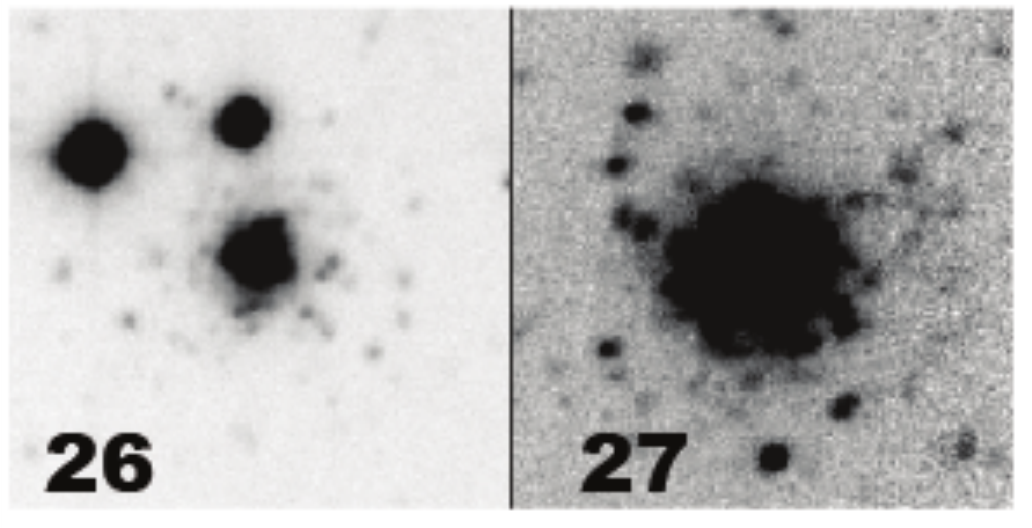}
 \vspace{2pt}
 \caption[g-band images of the new clusters (H25 and H26) from the Megacam data]
 {g-band images of the new clusters from the Megacam data.  Each image
   is also 30 x 30 arcsec.}\label{Fi:classical-mega}
\end{figure}

Our search for M31 halo clusters exploits the fact that GCs at this
distance should be just resolved in good seeing. In fact, given the
survey depths, one should even be able to resolve stars to below the
tip of the red giant branch (TRGB) in M31 clusters, enabling direct
detection of the brightest individual red giants. Confining the search
to the halo also means that the patchy extinction and rapidly
fluctuating light levels in the main body of M31 are not a concern.
For the INT/WFC survey, GC candidate selection was based on magnitude,
colour, ellipticity, object classification (stellar/non-stellar) and
image width (scale size), all of which are provided by the photometric
pipeline. GCs are known to inhabit a specific range of absolute
magnitudes, $-10.5 < V < -3.5$ (equivalent to apparent V magnitudes of
$14 < V < 21$ at the distance of M31), and colours, $0 < (V-I) <
1.7$.  The generous values, for both
magnitude and colour, were chosen so as to allow for errors.  This
conservative approach was chosen to avoid excluding any possible GCs
from the candidate list generated (but even these parameters finally
proved over-constraining, see the following section).  Ellipticity was
used as a criterion ($ < 0.3$) to eliminate many field galaxies, and
finally the image width (having a Gaussian $\sigma > 2.5$ pixels, which for the
INT/WFC is equivalent to $\approx$ 0.85 arcsec - comparable to the
best seeing in the survey images) of objects was employed to remove
the bulk of stars. These cuts were tested against the then confirmed
M31 GCs to ensure that none were excluded\footnote{We are
grateful to the anonymous referee for pointing out that GCs with a ellipticities $>$ 0.3
have been found in NGC 5128 [\citealt{Harrisetal06}]. This result was published after our search
had been undertaken, and it is possible that very elliptical halo M31 GCs may await discovery.}
For the CFHT/Megacam
survey, the search was based on magnitude and colour cuts alone as the
greater resolution of Megacam allows better visual identification of
GC candidates. The same magnitude limits were used for the g and i band
employed by CFHT/Megacam as  the conservative V and I values allowed 
for the conversion to g and i. All GC candidates were then visually inspected for
morphological evidence of cluster status and, if confirmed as a
cluster, cross-checked against the Revised Bologna Catalogue
(http://www.bo.astro.it/M31/; \citealt{Galletietal04a, Galletietal06,
  Galletietal07}) to determine if they were previously known. Our
search has resulted in the discovery of 27 new ``compact'' GCs in the
halo of M31.  The V and g images of these clusters are shown in
Figures \ref{Fi:classical-int} and \ref{Fi:classical-mega} and their
basic positional and photometric properties are listed in Table
\ref{tab:compacts} (note that the H prefix indicates ``halo'' GC).
Nine of these new clusters were initially presented in a short
conference paper \citep{Huxoretal04}.  This sample included one
cluster (H6) which was subsequently rediscovered by
\citet{Galletietal05} who gave it an RBC designation of
B514. \cite{Mackeyetal07} have recently presented deep HST/ACS imagery for a
subsample of these classical GCs which they use to derive metallicities, structural
parameters and horizontal branch morphologies; the GC IDs used in
that paper are listed in column 6 of Table \ref{tab:compacts}.

We note that much recent work on extragalactic GCs has preferred the
use of colours that are more sensitive to metallicity, such as (B-V)
and (B-I) over (V-I), or variants on the Washington system
(e.g. \citealt{Dirschetal03}), and (V-K) optical/NIR colours
(e.g. \citealt{Puziaetal02}).  These are the bands of choice when the
aim is to obtain age and metallicity measures.  In the study presented
here, we are constrained to the existing V- (or g-) and i-band images
that have been taken as part of a larger general-purpose
survey. However, the failings of (V-I) over these other systems is
less important, as our aim here is simply to find candidates for
follow-up observations.

\begin{table}
 \centering
 \begin{minipage}{80mm}
  \caption{Properties of the New Classical Globular Clusters. The V magnitude was obtained for
  a eight arcsec aperture radius, the (V - I) colour for a smaller six arcsec aperture.}\label{tab:compacts}
  \vspace{2pt}
  \begin{tabular}{@{}lllccc@{}}
  \hline
   ID & RA(J2000) & Dec(J2000) & V &  (V-I)   & M07 ID\footnote{IDs used by \citet{Mackeyetal07}.}\\

 \hline
 H1  & 00 26 47.80 & +39 44 46.7 & 16.05 & 0.98  & GC1 \\
 H2  & 00 28 03.28 & +40 02 56.2 & 17.58 & 0.94  & \\
 H3  & 00 29 30.15 & +41 50 32.1 & 18.13 & 1.16  & \\
 H4  & 00 29 45.01 & +41 13 09.6 & 16.98 & 0.94  & GC2 \\
 H5  & 00 30 27.30 & +41 36 20.0 & 16.31 & 0.95  & GC3 \\
 H6  & 00 31 09.85 & +37 54 00.4 & 15.76 & 1.08  & GC4 \\
 H7  & 00 31 54.55 & +40 06 47.8 & 18.10  & 1.09  &  \\
 H8  & 00 34 15.44 & +39 52 53.2 & 19.62 & 1.24  & \\
 H9  & 00 34 17.29 & +37 30 43.3 & 17.62 & 0.94  & \\ 
 H10 & 00 35 59.76 & +35 41 03.9 & 16.09 & 1.08  & GC5 \\                         
 H11 & 00 37 28.12 & +44 11 25.0 & 16.88 & 0.96  & \\
 H12 & 00 38 03.87 & +37 44 00.7 & 16.47 & 0.97  & \\
 H13 & 00 38 33.65 & +41 44 53.1 &     - & -   \footnote{Cluster overlapped by bright star making photometry impossible}  & \\
 H14 & 00 38 49.39 & +42 22 47.1 & 18.27 & 1.20  & GC7 \\
 H15 & 00 40 13.20 & +35 52 36.6 & 17.99 & 0.84  & \\
 H16 & 00 40 37.80 & +39 45 29.9 & 17.53 & 0.92  & \\
 H17 & 00 42 23.68 & +37 14 35.0 & 17.37 & 0.97  & \\
 H18 & 00 43 36.03 & +44 58 59.3 & 16.64 & 0.98  & \\
 H19 & 00 44 14.88 & +38 25 42.2 & 17.35 & 0.96  & \\
 H20 & 00 45 52.51 & +39 55 52.3 & 18.63 & 0.85  &  \\
 H21 & 00 48 50.57 & +41 08 36.5 & 18.69 & 1.06  & \\
 H22 & 00 49 44.69 & +38 18 37.4 & 17.03 & 0.99  &\\
 H23 & 00 54 25.02 & +39 42 55.3 & 16.72& 1.04  & GC8 \\
 H24 & 00 55 43.94 & +42 46 16.2 & 17.78 & 1.07  & GC9 \\
 H25 & 00 59 34.56 & +44 05 39.1 & 17.29 & 1.47  & \\
 H26 & 00 59 27.37 & +37 41 34.1 & 16.96 & 0.98  & \\
 H27 & 01 07 26.30 & +35 46 46.5 & 16.50 & 0.91  & GC10 \\

\hline
\end{tabular}
\end{minipage}
\end{table}

\subsection{The Strategy for Extended Clusters}

 \begin{figure}
 \centering
 \includegraphics[angle=0,width=80mm]{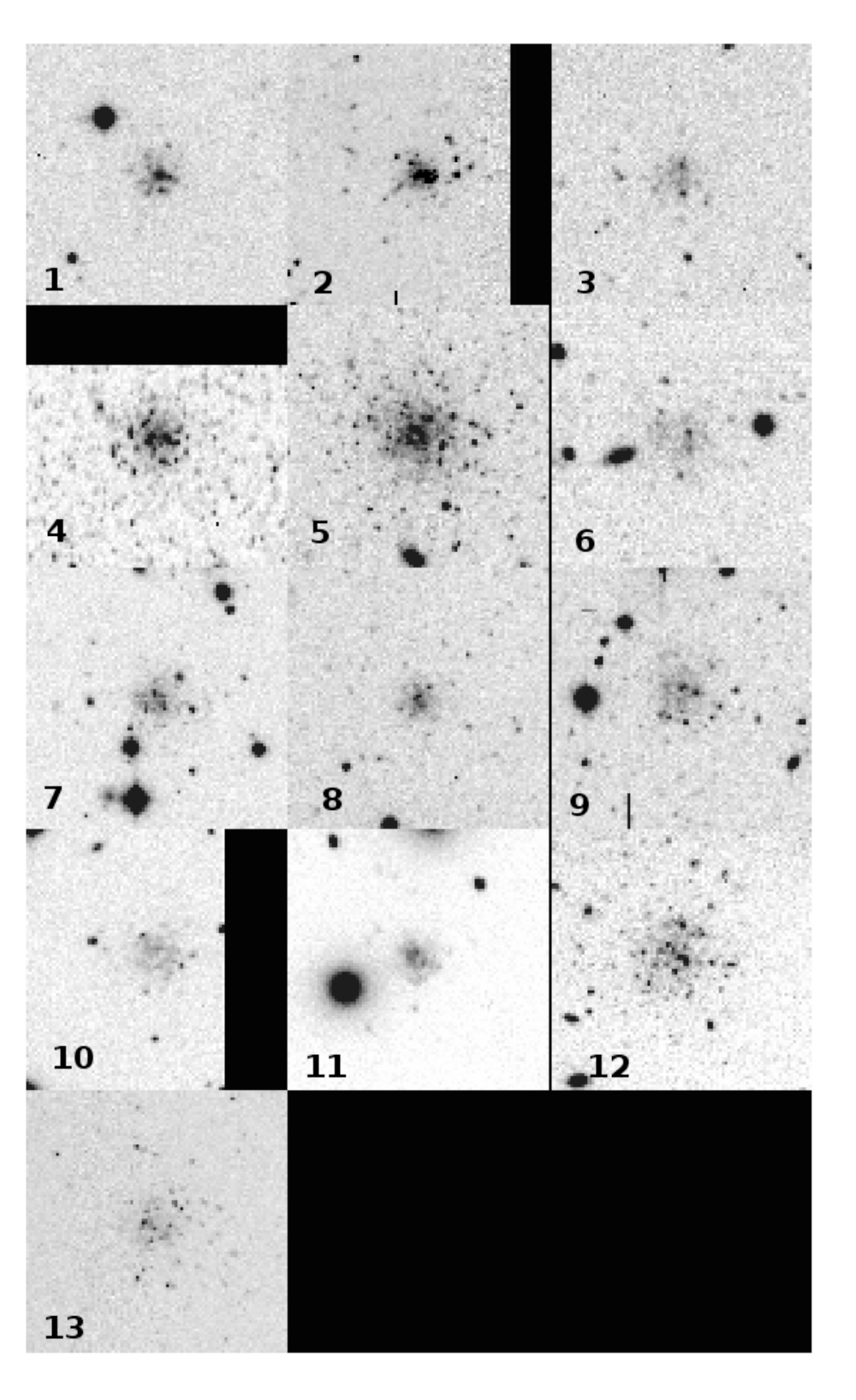}
 \vspace{2pt}
 \caption[V- or g-band images of the extended clusters]{V band (g band
   for clusters 2, 12 and 13, which are only present in the
   CHFT/Megacam survey area) images of the extended clusters. Fields
   are 1 arcmin $\times$ 1 arcmin. The numbers indicate the
   ID.}\label{Fi:fuzzies}
\end{figure}

While undertaking aperture photometry of one of the newly-discovered
classical GCs, a diffuse cluster was serendipitously discovered nearby
in the same field.  Since this object had been partially deblended by
the INT/WFC pipeline, it had been missed by the semi-automated
approach employed in the main GC survey. This object was significantly
more extended than typical GCs with a half-light radius ($R_{h}$) of
35 pc .  This discovery subsequently prompted a
visual survey of all the INT/WFC and CFHT/Meegacam images to be
undertaken, which resulted in the discovery of 13 further objects with
similar ``extended'' properties (see Figure \ref{Fi:fuzzies}).
\cite{Huxoretal05} presented the first extended clusters found in M31
while \cite{Mackeyetal06} have presented deep HST/ACS CMDs for four
such systems.

\begin{table}
 \centering
 \begin{minipage}{80mm}
  \caption{Basic Properties of the New Extended Clusters}\label{tab:fuzzies}
  \vspace{2pt}
  \begin{tabular}{@{}lllccc@{}}
  \hline
   ID & RA(J2000) & Dec(J2000) & V & (V-I) & M06 ID\footnote{IDs used by \citet{Mackeyetal06}.} \\
 
 \hline 
 HEC1 & 00 25 33.96 & +40 43 39.4 & 18.44 & 0.49   &  \\
 HEC2 & 00 28 31.65 & +37 31 24.4 & 19.17 & 1.07 & \\ 
 HEC3 & 00 36 31.72 & +44 44 16.7 & 19.62 & 1.10   &   \\
 HEC4  & 00 38 04.60  & +40 44 39.0 & 17.60  & 1.02  & EC3  \\
 HEC5  & 00 38 19.50  & +41 47 15.0   & 17.60  & 0.88  & EC1  \\
 HEC6  & 00 38 35.60 & +44 16 49.0 & 18.99 & 1.05  &  \\
 HEC7  & 00 42 55.00  & +43 57 28.0   & 17.10  & 0.93   & EC2 \\
 HEC8  & 00 45 26.90 & +40 13 47.0 & 19.19 & 1.28  &    \\
 HEC9  & 00 50 45.88 & +41 41 34.6 & 18.65 & 1.04   & \\
 HEC10 & 00 54 36.40 & +44 58 44.2 & 18.77 & 1.00  &   \\
 HEC11  & 00 55 17.32 & +38 51 02.5 & 17.69 & 1.00    &   \\
 HEC12 & 00 58 15.42 & +38 03 02.0 & 18.84 & 1.07   & EC4 \\
 HEC13  & 00 58 16.93 & +37 13 50.9 & 19.60 & 0.81   & \\
 
\hline
\end{tabular}
\end{minipage}
\end{table}

It is interesting to speculate as to why these extended clusters were
not uncovered in previous GC searches of M31. Most of the earlier
surveys concentrated on the main disk area of M31, and the high
surface brightness background and stellar crowding would make the
discovery of such faint and extended objects difficult. These problems
would not be present in halo searches, however the sensitivity and
spatial extent of previous surveys (many of them based on photographic
plates) would neither reveal them nor their nature. Three extended clusters from
this sample have already been published \citep{Huxoretal05}, along with King model profile fits. These
(HEC5, HEC7 and HEC4) are the  most luminous of the extended clusters, yet only have  V-band surface brightnesses within the half-light radius of 23.9, 22.6 and 24.0 mags arcsec$^{-1}$ respectively, making them hard to distinguish from a background low-surface brightness galaxy. 

The basic properties of the extended clusters are listed in Table
\ref{tab:fuzzies}. Being more extended than the objects in Table \ref{tab:compacts},
the ID numbers have the prefix HEC to indicate that they are ``halo
extended cluster''.  The extended nature of these clusters makes them
very interesting; some of them are larger for their luminosity
than any known GCs and start to encroach on a region of $R_{h}$--M$_V$
parameter space normally inhabited by dwarf spheroidal galaxies . 

In total, we have found 40 new clusters in the halo of M31. Using the
latest Revised Bologna Catalogue (taking the \citet{Galletietal06}
revisions and those in Table \ref{tab:updates} into account), the new
clusters represent an increase of more than $10\%$ of the total
confirmed sample, and an increase of more than a $75\%$ of those known
beyond 60\arcmin ($\approx 14$~kpc).  This very significant
improvement in the sample of clusters at large radius enables a much better 
analysis of the features of the halo GC system, and this will be
described a follow-up paper (Huxor et al. in prep).

\begin{figure}
 \centering
 \includegraphics[angle=-0,width=90mm]{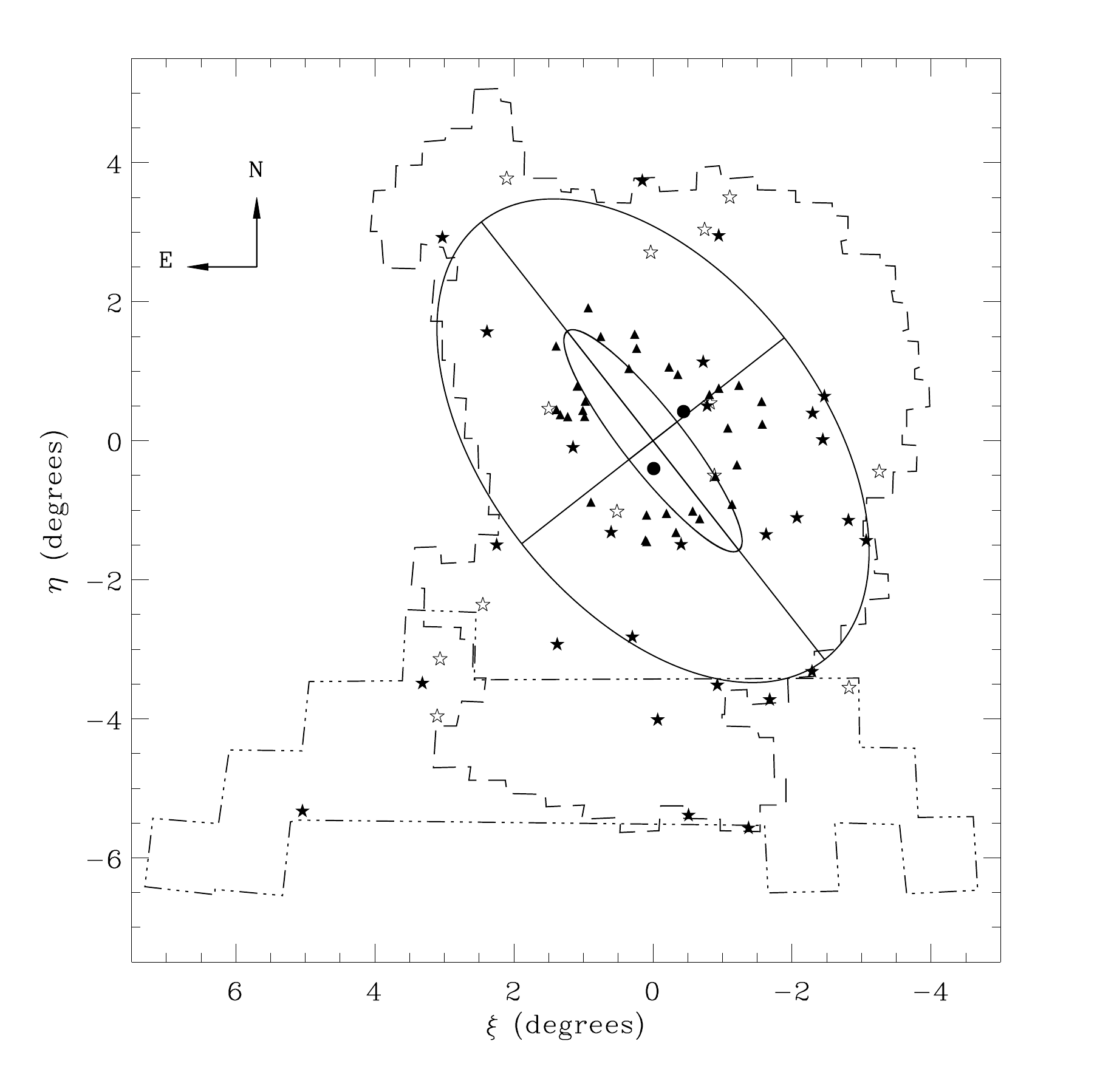}
 \vspace{2pt}
 \caption {The location of the new globular clusters (filled stars)
   and extended clusters (open stars) in relation to confirmed RBC GCs
   that lie in our survey area (black triangles; the remaining RBC GCs
   are not shown), to show the limited extent of the known GC
   population prior to this study.  The major landmarks (the satellite
   galaxies M32 and NGC 205 are also shown (large filled circles). The
   dashed line outlines the INT survey area covered, and the
   dashed-dotted line outlines that part of the Megacam survey
   employed.  The inner ellipse has a semimajor axis of 2$^\circ$ (27
   kpc) representing a disk with an inclination of 77.5$^\circ$; the
   optical disk of M31 lies well within this boundary.  The outer
   ellipse denotes a flattened ellipsoid of semimajor axis length
   4$^\circ$ (55 kpc).  Note that the relatively large size of the WFC fields result
   in a number of known GCs within the inner ellipse being found in our survey area.}\label{Fi:plot}
\end{figure}

We note that the new clusters -- both the classical and extended types
-- are found very far from M31 (see Figure \ref{Fi:plot}). They extend
far beyond the limits of previously ``confirmed'' GCs or the
spectroscopically confirmed sample of \cite{Perrettetal02}.  Indeed,
they are found in all regions that have been surveyed to date (up to $\sim$ 100 kpc),
suggesting that yet more are likely to be discovered as the survey
areas are extended in future.

\subsection{Photometic Properties}
\subsubsection{Optical Photometry}

\begin{figure}[h]
\begin{center}$
\begin{array}{cc}
\includegraphics[angle=0,width=80mm]{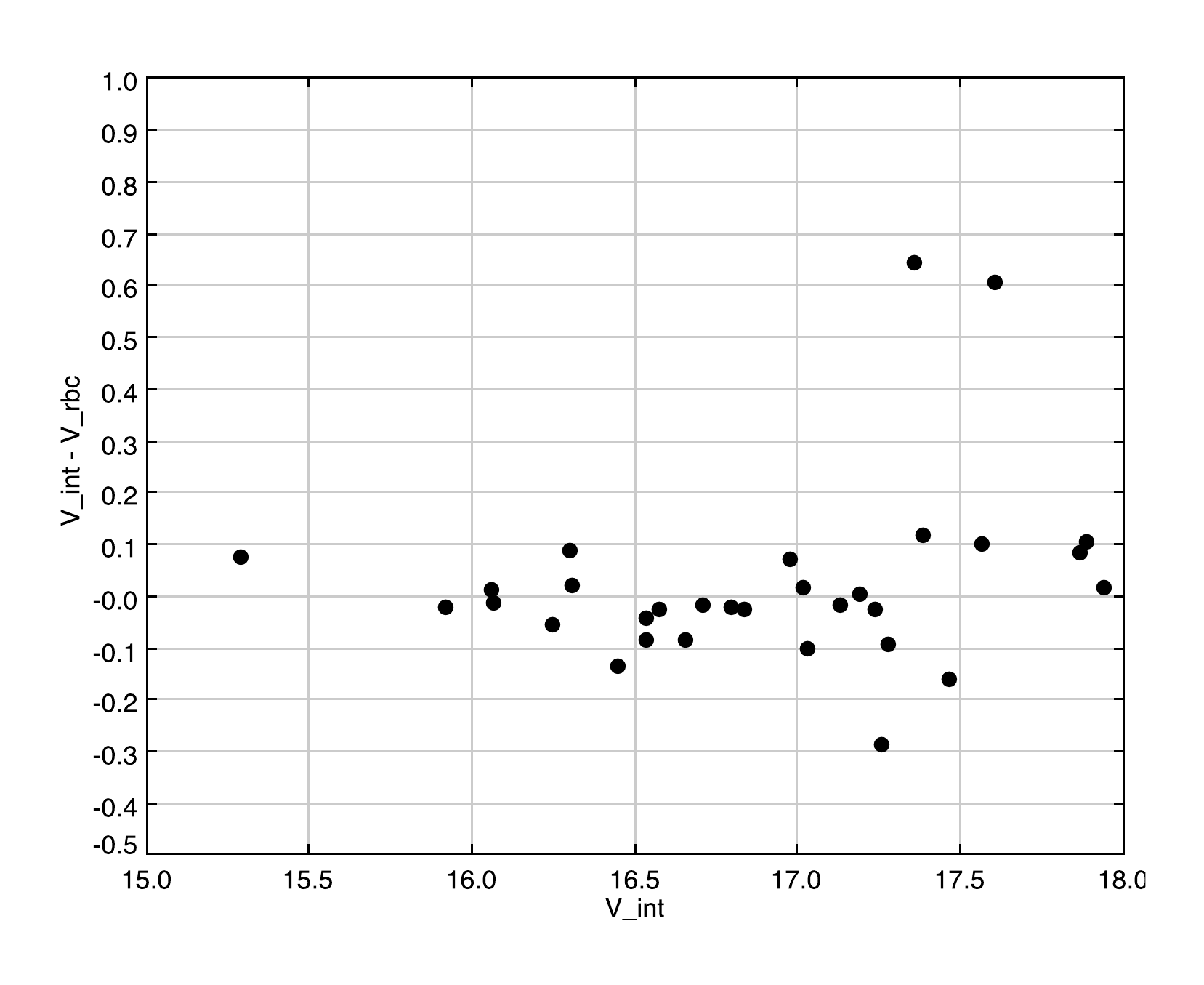} \\
\includegraphics[angle=0,width=80mm]{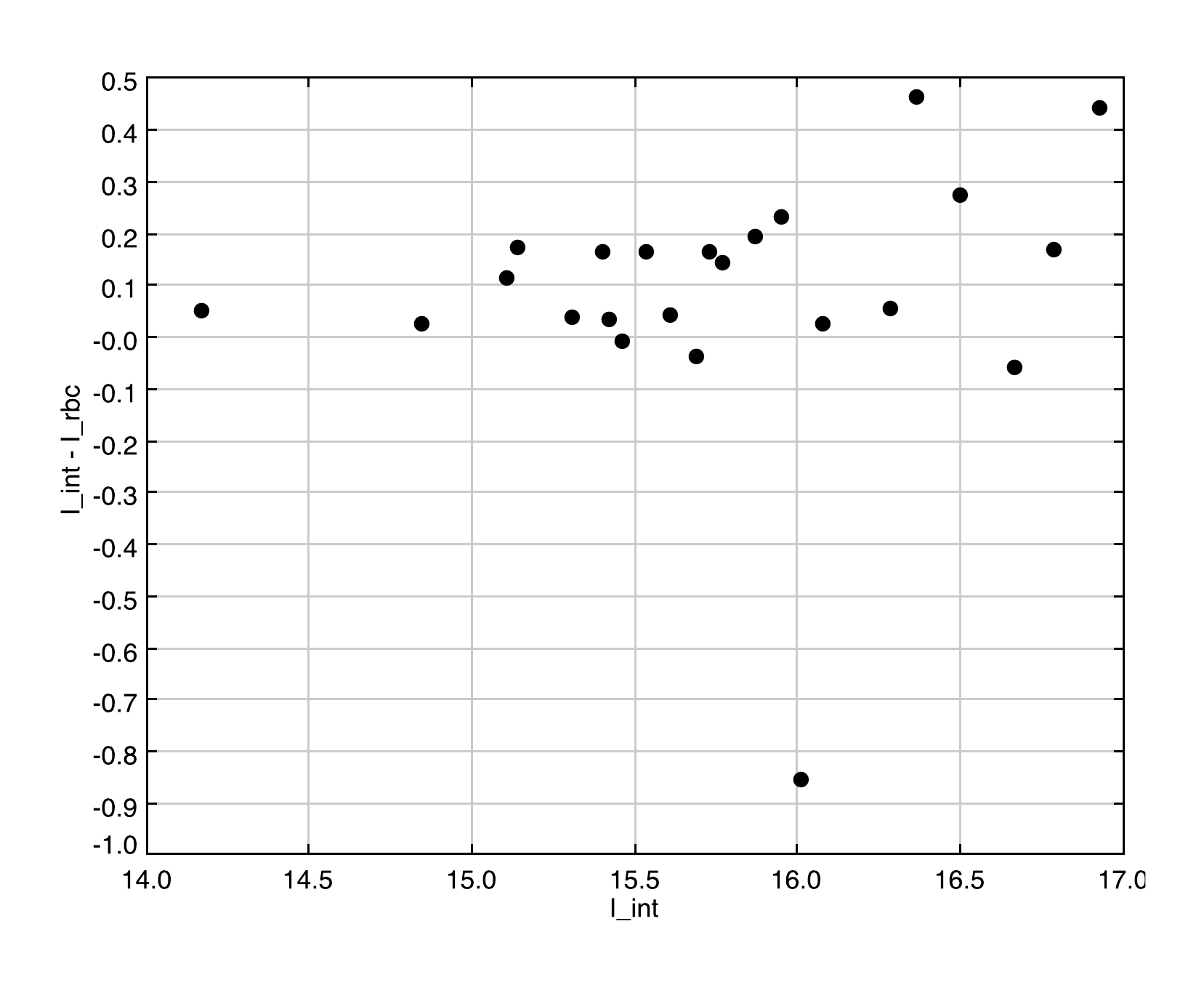} \\
\includegraphics[angle=0,width=80mm]{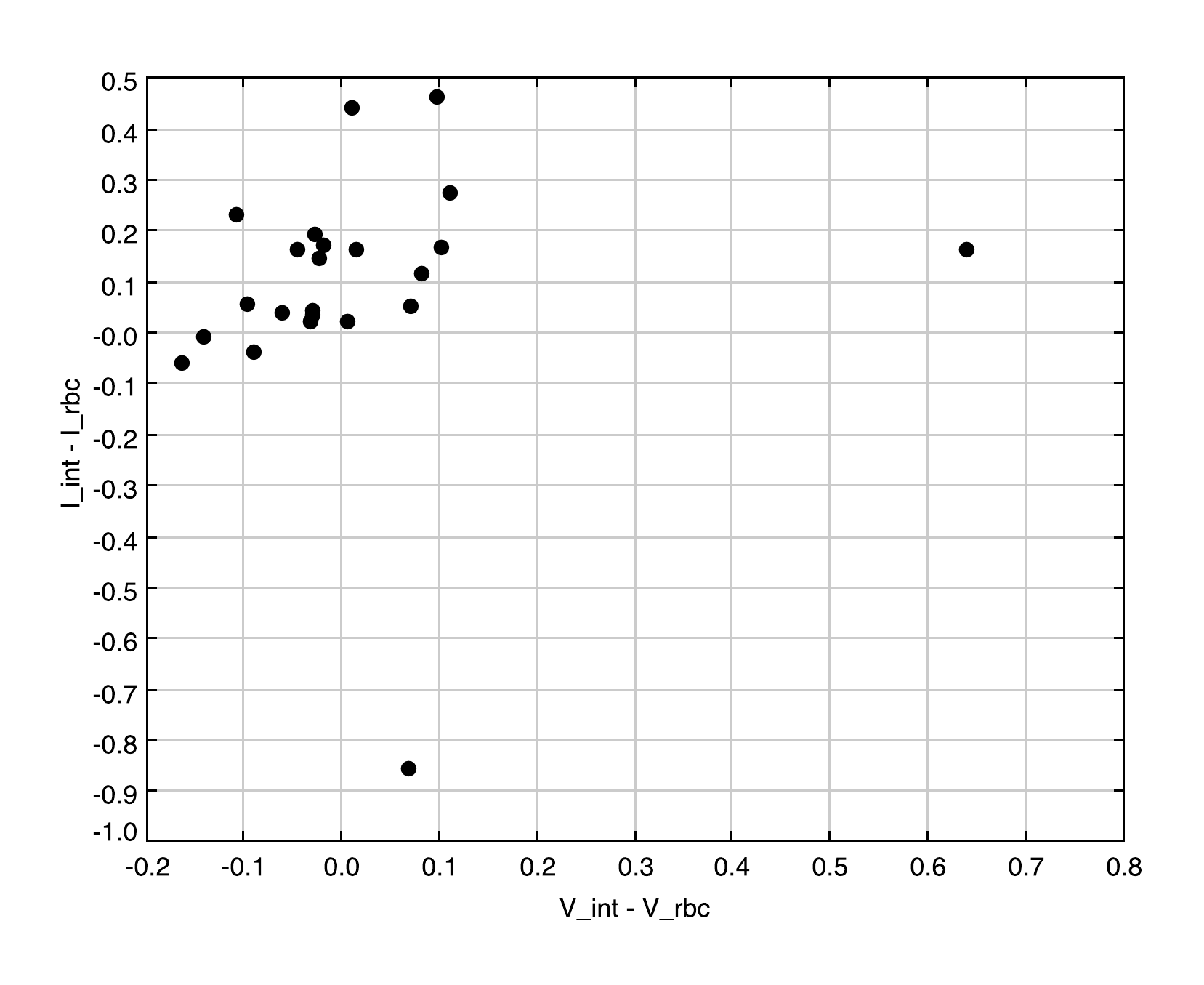} \\

\end{array}$
\end{center}
\caption{Residuals between new INT V and I band photometry, and published
	RBC values, and (lower panel) plot of v residuals against i residuals.}\label{Fi:phot_errors}
\end{figure}

Integrated photometry of the new GCs was undertaken with the IRAF task
apphot.  Table \ref{tab:compacts} lists the uncorrected Johnson V
magnitudes which have been measured within an aperture of 8\arcsec
radius except for clusters H12 and H16. The former is merged with a
bright star and has no photometry, while the latter has a value
determined at the edge of a CCD frame. The image for cluster H7 was
first masked to remove the spike from a nearby bright star. The
8\arcsec aperture was selected to allow direct comparison with the
photometry of \citet{Barmbyetal00}. Table \ref{tab:compacts} also
gives (V-I) within a 6\arcsec radius aperture. This smaller aperture
was used to reduce the error from the background and is valid assuming
that no colour gradient is present in the clusters, For the extended
clusters, the magnitudes were determined within a larger aperture of
12 arcsec radius which was judged to enclose the bulk of the light.

\begin{figure}
 \centering
 \includegraphics[angle=-0,width=90mm]{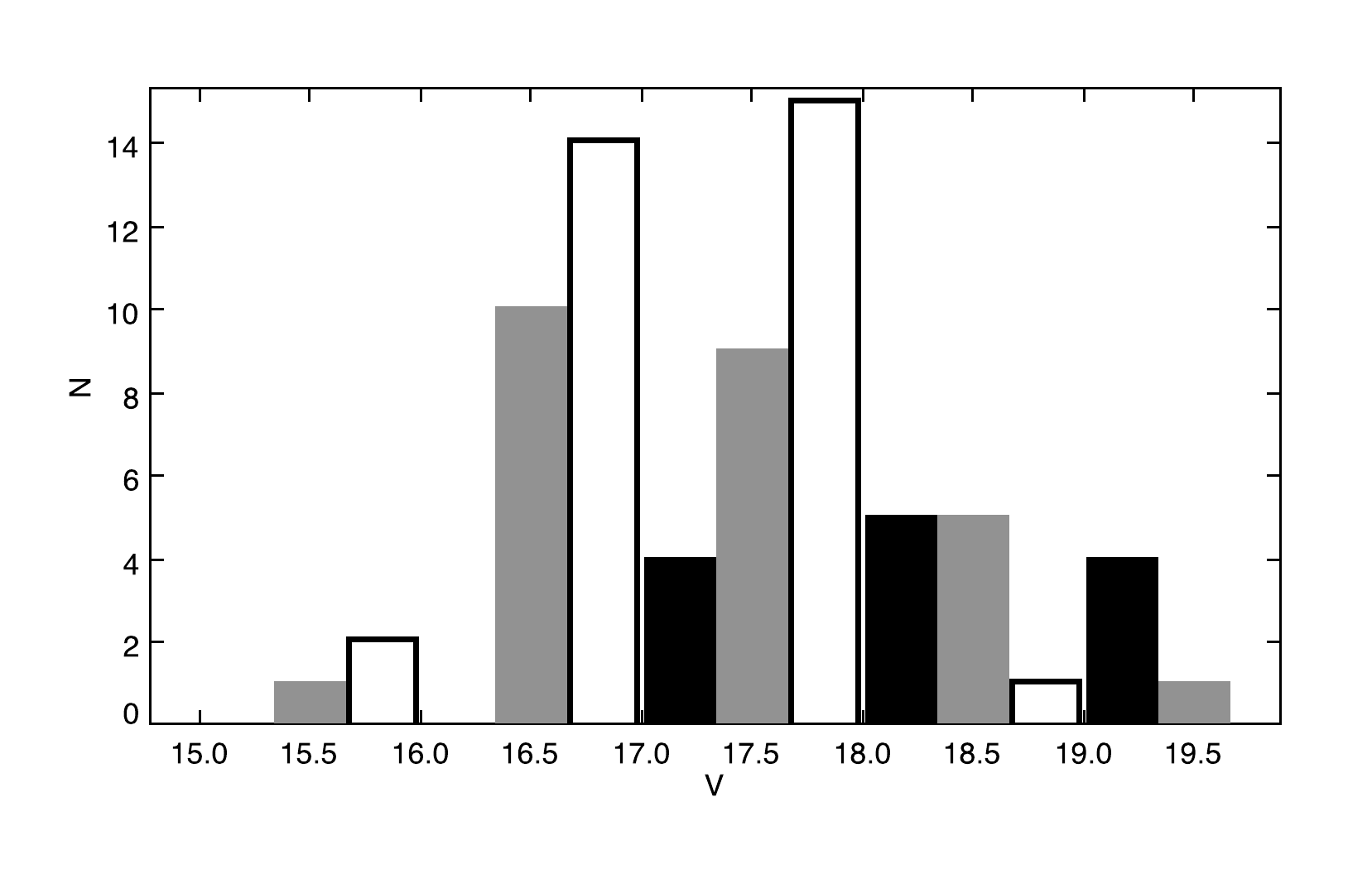}
 \vspace{2pt}
 \caption {Histograms of apparent magnitude for confirmed RBC GCs in
   our survey area (open bars), and the newly found compact (grey
   bars) and extended (black bars) GCs. }\label{Fi:lfs}
\end{figure}

The INT/WFC survey uses Johnson V (V$^{\prime}$) and Gunn i passbands
hence a colour transformation was required to obtain the Johnson/Cousins
equivalents V and I. This was done to allow for comparison of the GC
photometry with other work.  The transformations used are I = i -
0.101(V-I) and V = V$^{\prime}$ - 0.005(V-I)
\citep{McConnachieetal03}.  The CFHT Megacam data was transformed into
the standard V and I system, from the Megacam g and i$^{\prime}$ via the
transformations: \\
g$_1$ = g + 0.092 \\
i$_1$ = i$^{\prime}$ - 0.401 \\
(g - i) = g$_1$ - i$_1$ \\
I =  (i$_1$  - 0.401) - (0.08 $\times$ (g - i)$^2$)  + 0.06 \\
V = g$_1$ - (0.42  $\times$ (g - i) ) + (0.04  $\times$ (g - i)$^2$) + 0.10 

Due to the high
luminosity of the clusters, the formal errors reported by apphot are
quite small. Indeed, the cluster magnitude uncertainties are dominated by the
zero-point errors of $\sim$0.02 mag; colour errors are estimated to be
$\sim$ 0.03 mag. This does not include errors due to contamination from foreground stars or background galaxies, as these are difficult to estimate. 

We are confident that we exceed the RBC for depth. A plot of the V magnitudes for the newly discovered compact and extended clusters (\ref{Fi:lfs}), shows that they reach fainter than the RBC GCs in the halo (beyond the visible disk).
We would, however, still fail to find analogues of the very low luminosity GCs in the MW. For example, AM 4 would have a V$\sim$ 22.9 at the distance of M31. 

\subsubsection{Near-IR Photometry}

Near-IR photometry has been shown to be useful to reduce the
age-metallicity degeneracy inherent in optical-only observations
\citep{Puziaetal02}.  The majority of the new clusters have no 2MASS
K-magnitudes and those that do are sufficiently faint that the 2MASS
errors are $>$ 0.15 mag. We thus embarked on a program to obtain
near-IR photometry for a subset of the newly discovered clusters which
we briefly summarize here. UKIRT/UFTI K-band observations were taken
in queue mode on the 8th and 9th August 2004 (programme ID: U/04A/166,
see Table \ref{tab:log} for details) using a JITTER\_SELF\_FLAT\
recipe. UFTI \citep{Rocheetal03} is an IR imager, from 1-2.5$\mu$m,
which has a field-of-view of only 92 arcsec across.  Thirteen clusters
were observed.

\begin{table}
 \centering
 \begin{minipage}{80mm}
  \caption{Observation Log for UKIRT/UFTI and  K-band magnitudes}\label{tab:log}
  \begin{tabular}{@{}lllr@{}}
  \hline
   Date & Target  & Exposures (secs)  & $K_{s}$ \\

 \hline
 8th Aug. 2004   & H2 & 5 x 36     & 15.12    \\    
			 & H5 & 5 x 24      & 14.14    \\
  			 & H6 & 5 x 24    & 13.50 \\   
			 & H7 & 5 x 36      & 15.41    \\  
			 & H8 & 5 x 180   & 15.66     \\
			 & H10 & 5 x 24   & 13.81      \\
			 &  H14 & 5 x 36   & 15.07  \\
			 & H19 & 5 x 36     & 15.24    \\
  			& H23 & 5 x 24        & 14.18  \\
  			 & H24 & 5 x 36    & 15.17     \\
 9th Aug. 2004   & H4 & 5 x 36      & 14.55   \\
		           & H21 & 5 x 36      & 15.74     \\
  		           & H22 & 5 x 36      & 14.75   \\
\hline
\end{tabular}
\end{minipage}
\end{table}

The data were reduced using standard ORAC-DR pipeline. In addition to
the target clusters, a NIR standard star (FS103) was observed (5 x 5
second exposures) at the start and end of both nights. Aperture
photometry, employing IRAF/apphot, and matching apertures with the
optical photometry (8 arcsec), gave K-band magnitudes for the target
clusters (see Table \ref{tab:log}).

\section{Revision of RBC}

\subsection{Checking of RBC candidate classifications}

Due to the limited extent and contaminated nature of the existing
catalogues, it was decided to use the INT/WFC and
CFHT/Megacam surveys of M31 not only to search for new GCs in the outer
halo, but to clarify the nature of many candidates from the earlier
catalogues. In addition, other available archival imaging data was
incorporated where available (e.g. the M31 POINT-AGAPE microlensing survey
data \citep{CalchiNovatietal05} and the HST, CFHT/12K and
Suprime-Cam archives). 

For each RBC candidate beyond within our survey area (i.e. excluding the
visible part of M31) we inspected all the imagery
available to us. Table \ref{tab:updates} notes those cases where this
proccess led to an unambiguous amendment to the RBC classification.
This table also lists the file names of the appropriate archive images
to facilitate follow-up. In some cases, the status of a GC as
confirmed is unchanged.  In these situations, entries in Table
\ref{tab:updates} are only given where the new data improves
significantly on that already available in the RBC. For example, if
only a spectroscopic confirmation is given in the RBC we will note if
new direct visual confirmation is available. This may seem unnecessary
but we point out that M31 GCs do not always have radial velocities
distinct from MW foreground objects hence morphological confirmation
is valuable.  Likewise, any instance where new imagery is of much
higher quality than that previously used is also listed.  Some of these
amendments were also published in \citet{Galletietal07} however
many revisions are reported here for the first time.  We note that a
number of ``confirmed'' halo GCs were found to be questionable in the
INT data, and have been down-graded in this table from ``confirmed" to
candidate GC status. In these cases, the ``confirmed" status came from
radial velocity data only, but have values that are consistent with MW
stars. They do not show clear GC morphology, and their flag in the
INT-WFC pipeline reports them as stellar.

\begin{table*}
 \begin{minipage}{150mm}
   \caption{Updates to the Revised Bologna Catalogue. The previous and
     revised classes are those used by the RBC: 1. confirmed GC,
     2. candidate GC, 4. galaxy, 6. star. In addition, a new class, 8,
     shows objects that are blends of stars and/or
     galaxies. Note that there is frequently additional images available for any GC. That listed below is the image
     used for visual inspection.}\label{tab:updates}
  \vspace{2pt}
  \begin{tabular}{@{}lllll@{}}
  \hline
    ID & previous & revised &  data source &comment   \\
      & class & class &      &   \\

 \hline

B029 	&	1	&	1	&	 STIS O8GOI7HJQ	&	   	\\
B045D 	&	1	&	1(?)	&	 ACS J92GA5NAQ	&	very compact, cluster?    	\\
B047D 	&	1	&	6	&	 ACS J92GB2HWQ  	&	star, but faint, diffuse cluster nearby \\
 & & & & at 00:42:10.87 +41:29:58.5 	\\
B053 	&	1	&	6	&	 ACS J92GA1CYQ 	&		\\
B054D 	&	1	&	6	&	 ACS J8VP04010 	&	   	\\
B067D 	&	2	&	1	&	 ACS J92GB3DNQ 	&	very small  	\\
B072D 	&	2	&	4	&	 ACS J92GB6ZLQ 	&	 most likely a background galaxy  	\\
B075 	&	1	&	1	&	  ACS J92GA5NBQ 	&	   previously only a spectroscopic confirmation  	\\
B080 	&	2	&	1	&	 ACS J92GA5NBQ 	&	 new confirmation  	\\
B091D 	&	1	&	1	&	    ACS J92GA7VMQ 	&	   previously only a spectroscopic confirmation  	\\
B097 	&	1	&	1	&	 ACS J92GA6ZIQ  	&	  	\\
B098 	&	1	&	1	&	 ACS J92GA0C6Q 	&	   previously only a spectroscopic confirmation  	\\
B099D 	&	2	&	4	&	 INT 	&	   	\\
B100 	&	1	&	1	&	  ACS J92GC3CSQ 	&	    	\\
B101 	&	1	&	1	&	  WFPC2 U92GD101M  	&	   previously only a spectroscopic confirmation 	\\
B111 	&	1	&	1	&	 ACS J92GAC6Q 	&	   previously only a spectroscopic confirmation    	\\
B116  	&	1	&	1	&	 ACS J92GA3DLQ  	&	   previously only a spectroscopic confirmation  	\\
B122  	&	1	&	1	&	  ACS J92GA3DLQ  	&	   previously only a spectroscopic confirmation 	\\
B125 	&	1	&	1	&	 ACS J92GC1FKQ  	&	   previously only a spectroscopic confirmation  	\\
B135 	&	1	&	1	&	 ACS J92GA7VMQ  	&	   previously only a spectroscopic confirmation  	\\
B137 	&	1	&	1	&	 ACS J92GA7VMQ 	&	   previously only a spectroscopic confirmation  	\\
B141 	&	1	&	1	&	 ACS J92GA7VMQ  	&	   previously only a spectroscopic confirmation  	\\
B149 	&	1	&	1	&	  ACS J92GB7VOQ 	&	   previously only a spectroscopic confirmation,  \\
& & & & partly hidden by chip gap  	\\
B157 	&	2	&	1	&	  ACS J92GC2XJQ 	&	 new confirmation   	\\
B161 	&	1	&	1	&	  ACS J92GC2XJQ 	&	   previously only a spectroscopic confirmation   	\\
B164 	&	1	&	1	&	 ACS J92GD2XMQ  	&	   previously only a spectroscopic confirmation  	\\
B165 	&	1	&	1	&	 ACS J92GD2XJQ  	&	   previously only a spectroscopic confirmation  	\\
B175D 	&	2	&	4	&	 INT 	&	 shows object as a galaxy  	\\
B180 	&	1	&	1	&	  ACS J92GC6D1Q  	&	   previously only a spectroscopic confirmation  	\\
B182 	&	1	&	1	&	  ACS J92GC6D1Q  	&	   previously only a spectroscopic confirmation 	\\
B202D 	&	2	&	4	&	 INT 	&	 a fine spiral galaxy  	\\
B227D 	&	2	&	4	&	 INT 	&	 image shows spiral galaxy  	\\
B229D 	&	2	&	6	&	  CFHT12K  Stream-7	&	   	\\
B234D 	&	2	&	4	&	 INT 	&	 a spiral galaxy  	\\
B266 	&	2	&	1	&	 ACS J92GB4E7Q 	&	 new confirmation   	\\
B272 	&	1	&	1	&	  ACS J92GA8VUQ 	&	   	\\
B289D 	&	1	&	2	&	 INT 	&	   	\\
B292D 	&	1	&	2	&	 INT  	&	   	\\
B298 	&	1	&	1	&	 ACS J96G11010 	&	  cluster GC6 in \citet{Mackeyetal06}    	\\
B301 	&	1	&	1	&	 CFHT12K  550666p 	&	   previously only a spectroscopic confirmation  	\\
B305 	&	1	&	1	&	  In CFHT12K  550666p  	&	   previously only a spectroscopic confirmation 	\\
B309 	&	1	&	1	&	 In CFHT12K  550666p 	&	   previously only a spectroscopic confirmation   	\\
B350 	&	1	&	1	&	 CFHT12K Stream-8 	&	   	\\
B357  	&	1	&	1	&	  CFHT12K Stream-8 	&	   previously only a spectroscopic confirmation 	\\
B379 	&	1	&	1	&	 ACS  J8F857010 	&	   	\\
B407 	&	1	&	1	&	 ACS  J8DB07010 	&	   	\\
B427 	&	2	&	4	&	 INT 	&	 shows a blend of 2/3 galaxies  	\\
B443 	&	1	&	8	&	 INT 	&	 blend of three objects  	\\
B451 	&	1	&	8	&	 INT-WFC 	&	 appears to show a double star  	\\
B453 	&	1	&	8	&	 INT 	&	 blend of two stars  	\\
B463 	&	2	&	4	&	 CFHT12K Stream-8  	&	   	\\
BA11	&	1	&	2	&	INT	&	galaxy?	\\
BH20 	&	2	&	6	&	 ACS J8F101MRQ 	&	    	\\
G003 	&	1	&	8	&	 INT 	&	 blend of three objects   	\\
NB29  	&	2	&	2	&	 ACS J8VP04010, 	&	 asterism?  	\\
V234 	&	2	&	1	&	 STIS O4XCJ4ZZQ  	&	    	\\
\hline
\end{tabular}
\end{minipage}
\end{table*}

As can be seen from Figure \ref{Fi:plot}, the INT-WFC survey contains
almost all the known\footnote{This study was completed prior to the publication of
  \citet{Kimetal07}, which are therefore  not included in the RBC clarification process described in this section.} halo GCs in the Revised Bologna Catalogue.  It is thus possible to compare the INT/WFC
photometry of the known GCs with that reported in the RBC. This
enables a check on the RBC photometry -- which has been drawn from a
variety of sources, including photographic plates -- and provides an
improved and consistent set of photometry for all the GCs in the outer
halo.

Photometry was obtained for all ``confirmed'' RBC GCs which lie
outside the visible disk and which fall within in the WFC-INT M31 halo
survey area using the procedures previously described.  Table
\ref{tab:RBC_GCs} presents the new photometry for these systems as
derived from the INT/WFC survey.  The depth of the exposures in the
INT/WFC survey (typically 900 seconds) meant that the most luminous
GCs were either at or close to saturation and there is no entry
provided in the table. In other cases, the GC fell within the gaps
between the four CCDs of the WFC.

\begin{table}
 \centering
 \begin{minipage}{70mm}
  \caption[New Photometry of Known RBC Clusters]{New Photometry of Known RBC Clusters }\label{tab:RBC_GCs}
    \vspace{2pt}
  \begin{tabular}{@{}lll@{}}
  \hline
   ID (from RBC)  &  V (mag) & (V-I) \\

 \hline

B150D	&	17.26	&	0.89	\\
B167D	&	17.87	&	0.93	\\
B289	&	16.07	&	0.93	\\
B290	&	17.03	&	1.08	\\
B291	&	16.54	&	1.00	\\
B293	&	16.31	&	0.91	\\
B295	&	16.66	&	0.97	\\
B298	&	16.45	&	0.99	\\
B314         &       17.47        &       0.80 \\
B337	&	16.71	&	0.94	\\
B339	&	16.84	&	1.23	\\
B343	&	16.25	&	0.94	\\
B344D	&	17.02	&	0.99	\\
B357	&	16.58	&	1.16	\\
B358	&	15.29	&	1.12	\\
B365	&	17.36	&	1.63	\\
B396	&	17.28	&	0.99	\\
B398	&	17.57	&	1.20	\\
B399	&	17.39	&	0.89	\\
B401	&	16.80  	&	0.93	\\
B402	&	17.24	&	1.16	\\
B403	&	16.30  	&	1.19	\\
B407	&	16.06	&	1.21	\\
B422	&	17.94	&	1.01	\\
B457	&	16.98	&	0.97	\\
B468	&	17.89	&	1.10	\\
DAO25      &      18.87        &       1.14 \\
G260	&	17.61	&	1.59	\\
G268	&	16.54	&	1.19	\\
G327	&	15.92	&	0.97	\\
G339	&	17.19	&	1.03	\\
G353	&	17.13	&	0.92	\\

\hline
\end{tabular}
\end{minipage}
\end{table}

A plot of the difference between the previously published RBC
photometry and the new INT/WFC photometry is shown in Figure
\ref{Fi:phot_errors}.  The V-band values show typical residuals of
$\pm$ 0. 1 mag, with two outliers. One of these is G260, which the
INT/WFC imaging shows has two nearby stars, and may explain the offset
from the RBC value.  The I band shows a consistent positive residual,
rising up to 0.5 mag for the faintest magnitudes, and with the
expected increase in scatter (and with one outlier, B457, which also
has an unlikely V-I value of 0.04 from the RBC).  However, the RBC
notes that most of its I band values are derived from scanned
photographic plates and hence such errors are not unexpected.  The
good match between the RBC and INT/WFC V band photometry gives us
confidence in the quality of the photometry of the new clusters.

\section{Summary}

In this paper, we have presented the discovery of 40 new globular
clusters found in the halo of M31, out to a galactocentric distance of $\sim$ 100 kpc.
 Some of these these have been
published by us elsewhere but in this paper we present all the new GCs
for the first time.  They considerably increase the number of known GC
in the far halo. Of the 40 new clusters, 13 are found to be of an
extended nature. The new clusters are found all the way to the edge of
the survey area, strongly suggesting that others await discovery much
further out. Indeed, a search for new dwarf galaxies in an extension
to the Megacam survey used here has already found a luminous GC that
is, to date, the most distant known from the center of M31
\citep{Martinetal06}.  A future paper will take these new GCs, and the
recent updates to the RBC, to investigate the properties of the whole
GC system of M31 and the implications for our understanding of the
history of system.

\section*{Acknowledgments}

APH and AMNF are supported by a Marie Curie Excellence Grant from the
European Commission under contract MCEXT-CT-2005-025869. NRT acknowledges
a STFC Senior Research Fellowship. APH would
also like to thank the Centre for Astrophysics Research at the
University of Hertfordshire, who provided the PPARC PhD studentship
during which this work was undertaken and, more recently, provided
facilities to assist in the writing of this paper.  The Isaac Newton
Telescope is operated on the island of La Palma by the Isaac Newton
Group in the Spanish Observatorio del Roque de los Muchachos of the
Instituto de Astrof'sica de Canarias. This research also used the
facilities of the Canadian Astronomy Data Centre operated by the
National Research Council of Canada with the support of the Canadian
Space Agency.  Based on observations obtained with MegaPrime/MegaCam,
a joint project of CFHT and CEA/DAPNIA, at the Canada-France-Hawaii
Telescope (CFHT) which is operated by the National Research Council
(NRC) of Canada, the Institute National des Sciences de l'Univers of
the Centre National de la Recherche Scientifique of France, and the
University of Hawaii.  This publication makes use of data products
from the Two Micron All Sky Survey, which is a joint project of the
University of Massachusetts and the Infrared Processing and Analysis
Center/California Institute of Technology, funded by the National
Aeronautics and Space Administration and the National Science
Foundation.  This research has made use of NASA's Astrophysics Data
System Bibliographic Services.

\end{document}